\documentclass[amsmath, amssymb,10pt,sort&compress]{article}
\usepackage{amsmath,amssymb,graphicx}
\usepackage[dvips]{epsfig}
\usepackage{graphicx}
\usepackage{caption}
\usepackage{subcaption}
\usepackage{hyperref}
\usepackage{float}
\usepackage{xcolor}
\usepackage[utf8]{inputenc}
\restylefloat{figure}
\input{epsf.sty}
\newcommand{\nc}{\newcommand}
\nc{\ba}{\begin{eqnarray}}
\nc{\ea}{\end{eqnarray}}
\newcommand\be{\begin{equation}}
\newcommand\ee{\end{equation}}

\begin{document}

\title{Non-trivial Extension of Starobinsky  Inflation}
\author{Salomeh Khoeini-Moghaddam\\\emph{skhoeini(AT)khu.ac.ir}\\Department of Astronomy and High Energy Physics,\\ Faculty of Physics, Kharazmi University, Tehran, Iran}

\maketitle

\begin{abstract}

We consider a non-canonic field in the context of Starobinsky inflation. We work in Einstein-frame. In this frame, the gravitational part of the action is equivalent to the Hilbert-Einstein action, plus a scalar field called scalaron.
 We investigate a model with a heavy scalaron trapped at the effective potential minimum, where its fluctuations are negligible. To be more explicit, we consider a Dirac-Born-Infeld (DBI) field, which is usually considered within the brane inflation context, as the non-canonic field.
Although, the DBI field governs inflation through implicit dependence on Scalaron the boost factor, and other quantities are different from the standard DBI model. For appropriate parameters, this model is consistent with the Planck results.

{\bf keywords}:Early universe; Inflation; F(R) theory; Starobinsky model; Dirac-Born-Infeld(DBI) inflation;Brane inflation, scalar-tensor, multi-field inflation

\end{abstract}

\section{Introduction}
Inflation theory is proposed to solve fundamental problems of standard cosmology\cite{Guth:1980zm, Linde1, Albrecht}; it also explains the origin of the primordial fluctuations. Although observational data support the inflation theory in general, there is no fundamental theory that can describe the nature of this theory. In the simplest model, the inflaton field, which is responsible for inflating the universe, rolls down in an almost flat potential (slow-roll regime).
Observational data indicate that in single field models, the monomial potentials are disfavored, including the famous potential $m^2\phi^2$. Other models with more intricate potentials, especially with exponential tails, provide good fits to data. Brane inflation is another model that is consistent with the Planck data\cite{Planck 2018a, Planck2018b}.
One example of brane inflation is $D3-\bar{D}3$ which is a well-motivated scenario\cite{brane1,brane2,brane3,brane4}. In this scenario, due to an attractive force, the $\bar{D}3$-brane is sitting down at the bottom of a warped throat, while the $D3$ -brane is relatively mobile. When the $D3$-brane and the $\bar{D}3$-brane collide and annihilate, inflation ends. When the branes start close to or inside the throat, we can approximate the potential with a simple expression\cite{shandera06}.

In general, inflation can be derived from non-canonical fields. This kind of model is investigated in k-inflation\cite{mukhanov1,mukhanov2} and general multi-field inflation context\cite{langlois2008,koyoma2008}. Dirac-Born-Infeld(DBI) model of inflation\cite{Alishahiha}, which is first considered in the context of brane-inflation, is one of these models. Under some constraints, the DBI model is consistent with observational data \cite{Planck2019,gomes2018}.

On the other hand, in recent years, $F(R)$ theories have attracted attention. These theories, which are an extension of Hilbert-Einstein's action, are another approach to explain the acceleration periods of our universe. Maybe the simplest and most famous model of $F(R)$ theory is $R+cR^2$. Being within the Planck $68\%$ confidence level constraints arouse enthusiasm for this model\cite{Planck2018b}.
This model is proposed many years ago by Starobinsky \cite{Staro1, Stao2} as a model for inflation. It is usual to write the action in the Jordan frame. If one transforms to the Einstein frame, the action is equivalent to a scalar field plus Hilbert-Einstein action \cite{witt}. This dual scalar field, which is called scalaron, can take the role of inflaton with exponential potential.

Inspired by string theory and high-energy physics, there is motivation to have more than one field. A multi-field model has more phenomenology than a single field model. Many works consider extra fields in $F(R)$ theory\cite{Bamba:2015uxa, Myrzakulov2015,Canko_2020,Gomes_2017,Mori17, Elizalde_2019, Antoniadis19}. The simplest extension of the Starobinsky model is considering extra-canonical scalar fields in $R+cR^2$ gravity. It is shown that these models, with minimal and non-minimal couplings, are robust models\cite{Bruck2015, Starobinsky-like two-field inflation}.
It is also possible to add fields with more phenomenology such as the Higgs field or fields inspired by super-gravity and other fundamental theories\cite{Ketov2020-2,Higgs_dependent_scalaron}. Recently Starobinsky inflation is explored within the context of supergravity \cite{Ketov2020-3,Ketov2020-1,Aldabergenovetal2020}. In\cite{dbisupergravity,Ketov2018-1,Ketov2018-3,Ketov-supergravity}
 DBI field and D-brane models are investigated in the extension of supergravity to Starobinsky inflation.

Therefore, the natural question that arises is whether adding a new field with a nontrivial kinetic term to the $R+cR^2$ model would make a robust model.
This work aims to investigate the existence of a DBI action in the context of $R+cR^2$ gravity. From another point of view, we would like to study the brane inflation in the context of Starobinsky gravity. Apart from the theoretical origin, the square root feature of DBI action makes several novelties. In principle, every $F(R)$ theory can be reformulated  as a scalar-tensor theory. Therefore the DBI field in the $R+cR^2$ gravity is equivalent to the DBI field, supplemented by another scalar field. In our case, the  scalar field is scalaron. When the scalaron traps at its minimum, the DBI field governs inflation. Especially when we have heavy scalaron, only the DBI field impact on cosmological perturbation, so we only consider the DBI field perturbations in the observational parameters such as spectral index.
As mentioned before, we transform into the Einstein frame and redefine the fields. This redefinition causes the DBI field to couple with the dual field. This coupling modifies the dynamics of the DBI field and hence affects the cosmological parameters.

This paper is organized as follows: in section (\ref{Setup}), the setup of the model is described. In section (\ref{DBI}), the background solution is considered. The field perturbations are investigated in section (\ref{perturbation}). We also do some numerical analysis in (\ref{numerical analysis}). We summarized our results in section (\ref{conclusion}).
\section{The Setup }\label{Setup}
In principle, a generic $F(R)$ model is given by the below action;
\ba
S=\frac{1}{2\kappa^2}\int d^4x\sqrt{-g}f\left(R\right)
\ea
This model is connected to the scalar-tensor theory via Legendre transformation as,
 \ba
 S=\frac{1}{2\kappa^2}\int d^4x\sqrt{-g}\left(f\left(\phi\right)+f'\left(\phi\right)\left(R-\phi\right)\right).
 \ea
Where we defined $\Omega^2\equiv f'\left(\phi\right)$ and $\phi$  is a real scalar field. With this definition, we rewrite the above action as,
\ba
S=\int d^4x\sqrt{-g}\left(\frac{1}{2\kappa^2}\Omega^2R-V\left(\phi\right)\right),
\ea
with $V\left(\phi\right)\equiv\frac{1}{2}\left(\phi f'\left(\phi\right)-f\left(\phi\right)\right)$.
The stability of classical and quantum gravity requires having $f'\left(R\right)>0$ and $f''\left(R\right)>0$ where $'$ denotes derivative with respect to $R$.

 It is also possible to add a matter sector. We are interested in matter with the non-canonic kinetic term; the general action can be written as below,
\ba
S=\int d^4x\sqrt{-g}\left(\frac{1}{2\kappa^2}\Omega^2R-V\left(\phi\right)\right)+\int d^4x\sqrt{-g}P\left(\mathbb{X}_\chi,\chi\right)
\ea
where $\chi$ is the non-canonic field and $\mathbb{X}_\chi=-\frac{1}{2}g_{\mu\nu}\partial_\mu\chi\partial_\nu\chi$ is its kinetic term.  $P$ denotes the Lagrangian density of the matter field; it is a function of both $\mathbb{X}$ and  $\chi$.

It is feasible  to go to  Einstein-frame under a conformal transformation $\tilde{g}_{\mu\nu}=\Omega^2g_{\mu\nu}$, we define a new field as $\Omega^2=f'\left(\phi\right)=e^{2\alpha\psi}$.
First, we consider the gravitational part of the action,
\ba\nonumber
S'_G&=&\int d^4x\sqrt{-\tilde{g}}\left(\frac{\tilde{R}}{2\kappa^2}-\frac{1}{2}\tilde{g}^{\mu\nu}\tilde{\partial}_\mu\psi\tilde{\partial}_\nu\psi-
\tilde{V}\left(\psi\right)\right),
\ea
where $\alpha=\frac{\kappa}{\sqrt{6}}$ with  $\kappa^2=8\pi G=M^{-2}_{pl}$. $M_{pl}$ is the reduced Planck mass. Under transformation to the Einstein frame, the matter part transforms as below,
\ba
S'_M=\int d^4x \sqrt{-\tilde{g}}e^{-4\alpha\psi}P\left(\mathbb{\tilde{X}}_\chi,\chi\right)
\ea
where $\mathbb{\tilde{X}}_\chi=-\frac{1}{2}\tilde{g}_{\mu\nu}\tilde{\partial}_\mu\chi\tilde{\partial}_\nu\chi$; $\tilde{\partial}$ indicates the derivative with respect to $\tilde{g}_{\mu\nu}$.
In the Einstein frame, there are two fields; the first one is scalaron, denoted by $\psi$. It comes from the correction of Einstein's gravity. The second one is  $\chi$, the matter field. Both of these fields influence inflation. The conformal transformation causes  $\chi$ to be coupled with $\psi$. In fact, F(R) theories are equivalent to scalar-tensor models\cite{Ketov2020-2,Maeda_book}; this equivalence permits us to apply the same  formalism to F(R) models.
In the next section, we focus on $R+cR^2$  and a special kind of non-canonic field i.e. Dirac-Born-Infeld(DBI) field.

\section{DBI field Dynamics in Starobinsky Model}\label{DBI}
In this section, we consider Starobinsky action. As mentioned before, the Starobinsky model is a robust model\cite{Robustness-Starobinsky}.
 We choose a DBI field as the non-canonic field,
\ba\label{action1}
S&=&\frac{1}{2\kappa^2}\int d^4x\sqrt{-g}(R+\mu R^2)\\\nonumber
&+&\int d^4x\sqrt{-g}[\frac{1}{f(\chi)}\left(1-\sqrt{1+f(\chi)g^{\mu\nu}\partial_\mu\chi\partial_\nu\chi}\right)-U\left(\chi\right)],
\ea
where $\kappa^2=8\pi G=M^{-2}_{pl}$.
In the following, we will work in natural units in which $\kappa^2=1$. The coupling parameter $\mu$, with units $[mass]^{-2}$, is assumed to satisfy the condition $\mu \gg \kappa^2$. In the following, we set $\mu\sim10^{9}M^{-2}_{pl}$, as it is fixed by the observed CMB amplitude. In the DBI part, $f\left(\chi\right)\approx\frac{\lambda}{\chi^4}$ is the warp factor of DBI field and $U(\chi)$ is its potential. Originally, this model proposed in the context of $D3-\bar{D}3$ brane-inflation in a warped throat. We assume D3-brane starts inside the throat, so the effective potential takes the simple form as\cite{shandera06},
\ba\label{chi-potential}
U\left(\chi\right)&=&\frac{1}{2}m^2\chi^2+V_0\left(1-\frac{vV_0}{4\pi^2}\frac{1}{\chi^4}\right)
\ea
 $V_0$ is the effective cosmological constant ; it depends on the warp factor of the $\bar{D}3$ branes position.  The constant v  depends on the properties of the warped throat,  we choose v = 27/16 ( see \cite{shandera06} and references therein).

 Therefore, The total action in Einstein-frame is given by,
\ba\label{full action}
S'&=&\int d^4x\sqrt{-\tilde{g}}\left(\frac{\tilde{R}}{2\kappa^2}-\frac{1}{2}\tilde{g}^{\mu\nu}\tilde{\partial}_\mu\psi\tilde{\partial}_\nu\psi-
e^{-4\alpha\psi}\frac{\left(e^{2\alpha\psi}-1\right)^2}{8\kappa^2\mu}\right)\\\nonumber
&+&\int d^4x\sqrt{-\tilde{g}}e^{-4\alpha\psi}\left(\frac{1}{f\left(\chi\right)}
\left(1-\sqrt{1+f\left(\chi\right)e^{2\alpha\psi}\tilde{g}^{\mu\nu}\tilde{\partial}_\mu\chi\tilde{\partial}_\nu\chi}\right)-U\left(\chi\right)\right).
\ea
From the above equation, we can read the potential of $\psi$ as
\ba\nonumber
w\left(\psi\right)\equiv \frac{1}{8\kappa^2\mu}e^{-4\alpha\psi}\left(e^{2\alpha\psi}-1\right)^2.
\ea
 The mass of scalaron is also defined as $m^2_\psi=\frac{1}{6\mu}$. We assume the metric of space-time is flat FRW, $ds^2=-dt^2+a^2\left(t\right)d\vec{x}^2$;
 then the equations of motion for $\psi$ and $\chi$  are as follows,
\ba\label{eqm1}
\ddot{\psi}+3H\dot{\psi}+w_{,\psi}=-\alpha e^{-4\alpha\psi}T^b_{DBI},
\ea
\ba\label{eqm2}
\ddot{\chi}&+&3H\gamma^{-2}\dot{\chi}+e^{-2\alpha\psi}\frac{f_{,\chi}}{2f^2}\left(1+2\gamma^{-3}-3\gamma^{-2}\right)+e^{-2\alpha\psi}\gamma^{-3}U_{,\chi}\\\nonumber
&=&\alpha\dot{\psi}\dot{\chi}\left(3\gamma^{-2}-1\right).
\ea
Where, $\gamma=1/\sqrt{1-e^{2\alpha\psi}f\dot{\chi}^2}$, is the modified boost factor of DBI field. The presence of $e^{2\alpha\psi}$ under the square root affects the dynamics of $\chi$. $T^b_{DBI}\equiv [f^{-1}\left(\chi\right)\left(4-\gamma-3\gamma^{-1}\right)-4U\left(\chi\right)]$ is the trace of the energy-momentum tensor of the DBI part. $()_{,\psi}$ and $()_{,\chi}$ denote derivative with respect to the fields $\psi$ and $\chi$, respectively.
  Einstein's field equations in flat FRW background are given as below,
\ba
3H^2&=&\frac{1}{2}\dot{\psi}^2+e^{-4\alpha\psi}\frac{\left(e^{2\alpha\psi}-1\right)^2}{8\kappa^2\mu}
+\rho_{DBI}\label{H}\\
-2\dot{H}&=&\dot{\psi}^2+e^{-2\alpha\psi}\dot{\chi}^2\gamma,\label{dotH}
\ea
where $\rho_{DBI}=e^{-4\alpha\psi}[f^{-1}\left(\gamma-1\right)+U\left(\chi\right)]$.

\begin{figure*}[t!]
 \begin{subfigure}[t]{0.5\textwidth}
  \includegraphics[height=3cm]{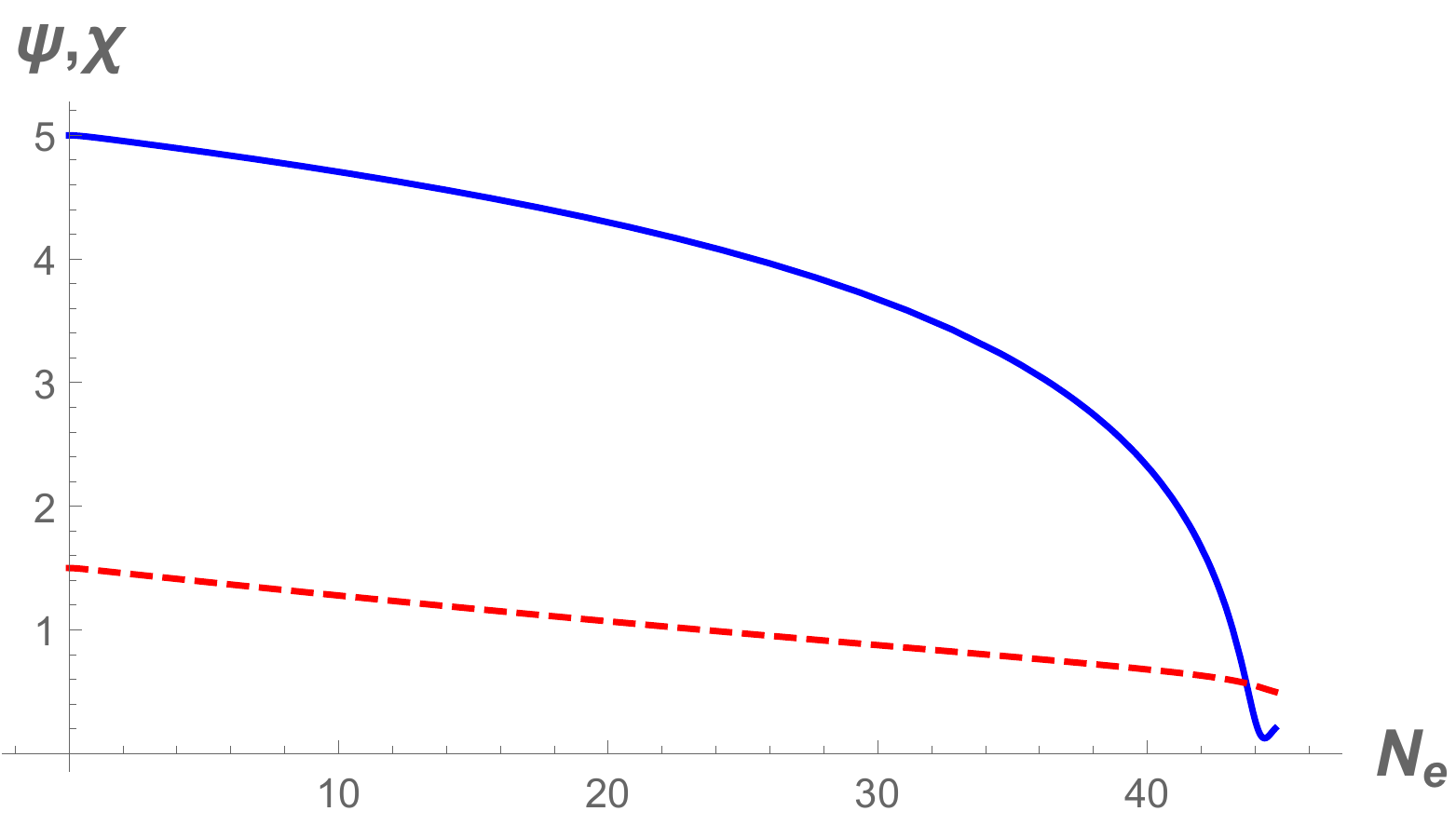}
   \caption{mass ratio=1}
 \end{subfigure}
 \begin{subfigure}[t]{0.5\textwidth}
    \includegraphics[height=3cm]{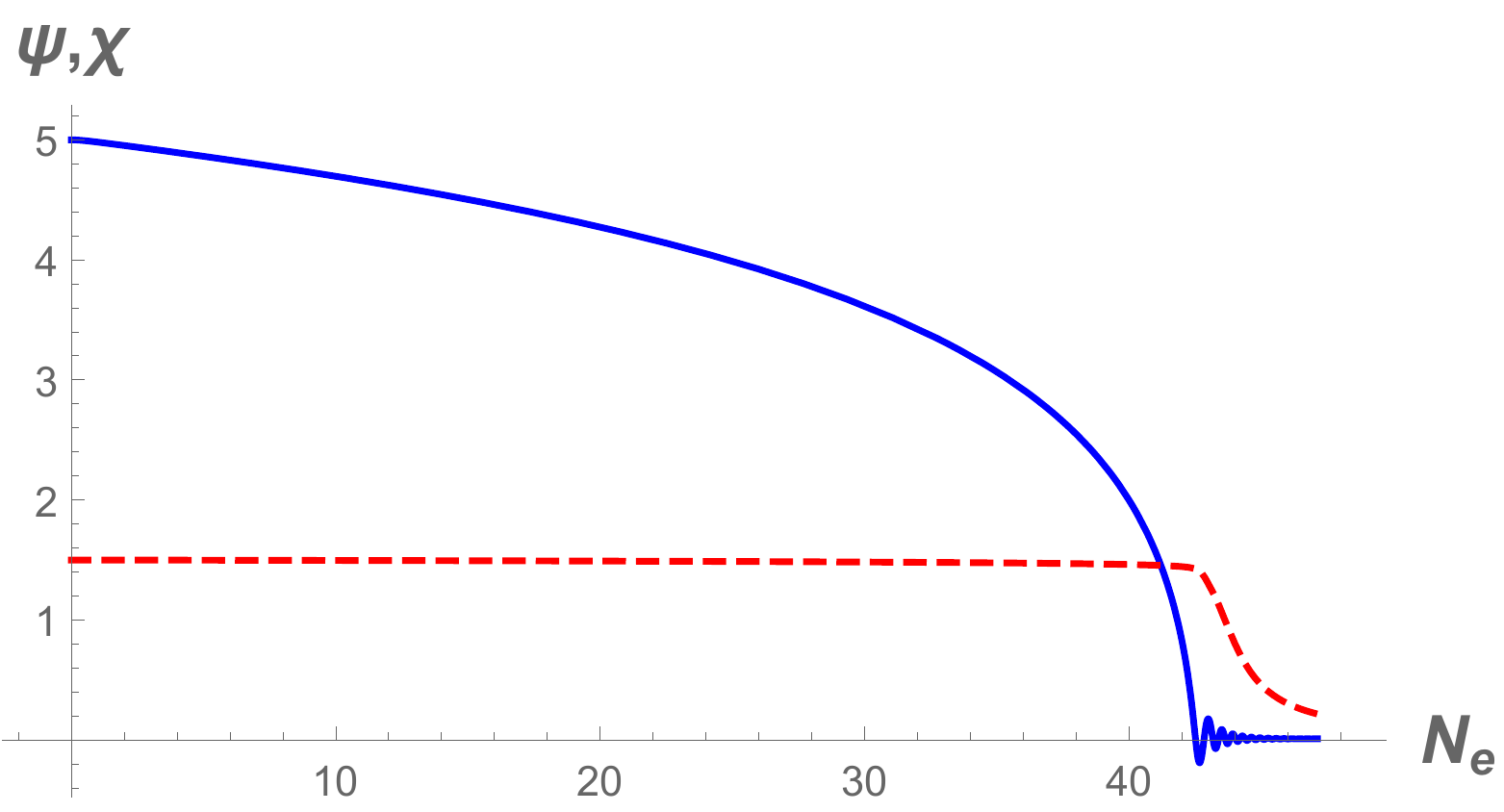}
      \caption{mass ratio=10}
 \end{subfigure}
 \begin{subfigure}[t]{0.5\textwidth}
  \includegraphics[height=3cm]{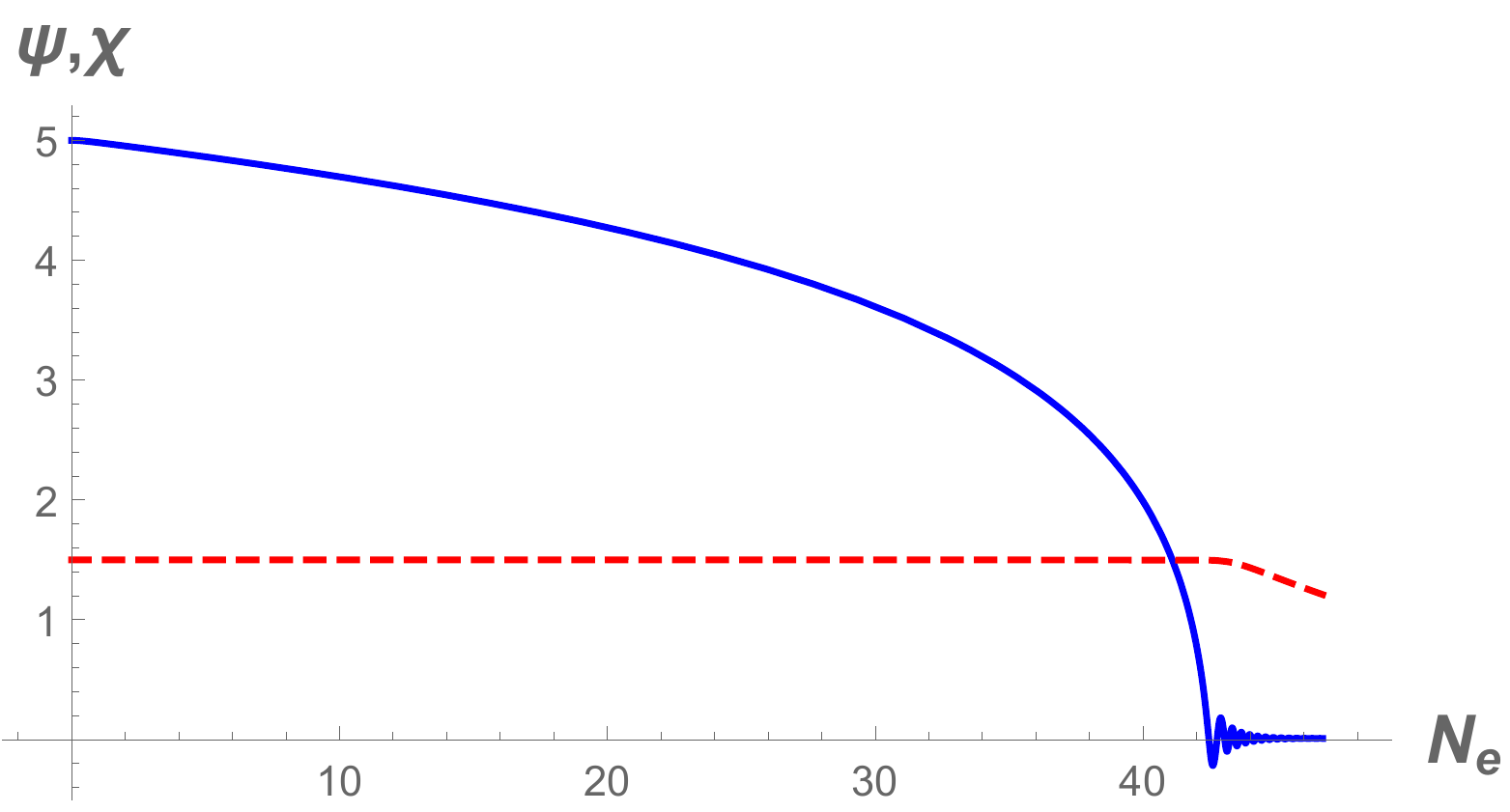}
   \caption{mass ratio=50}
 \end{subfigure}
 \begin{subfigure}[t]{0.5\textwidth}
  \includegraphics[height=3cm]{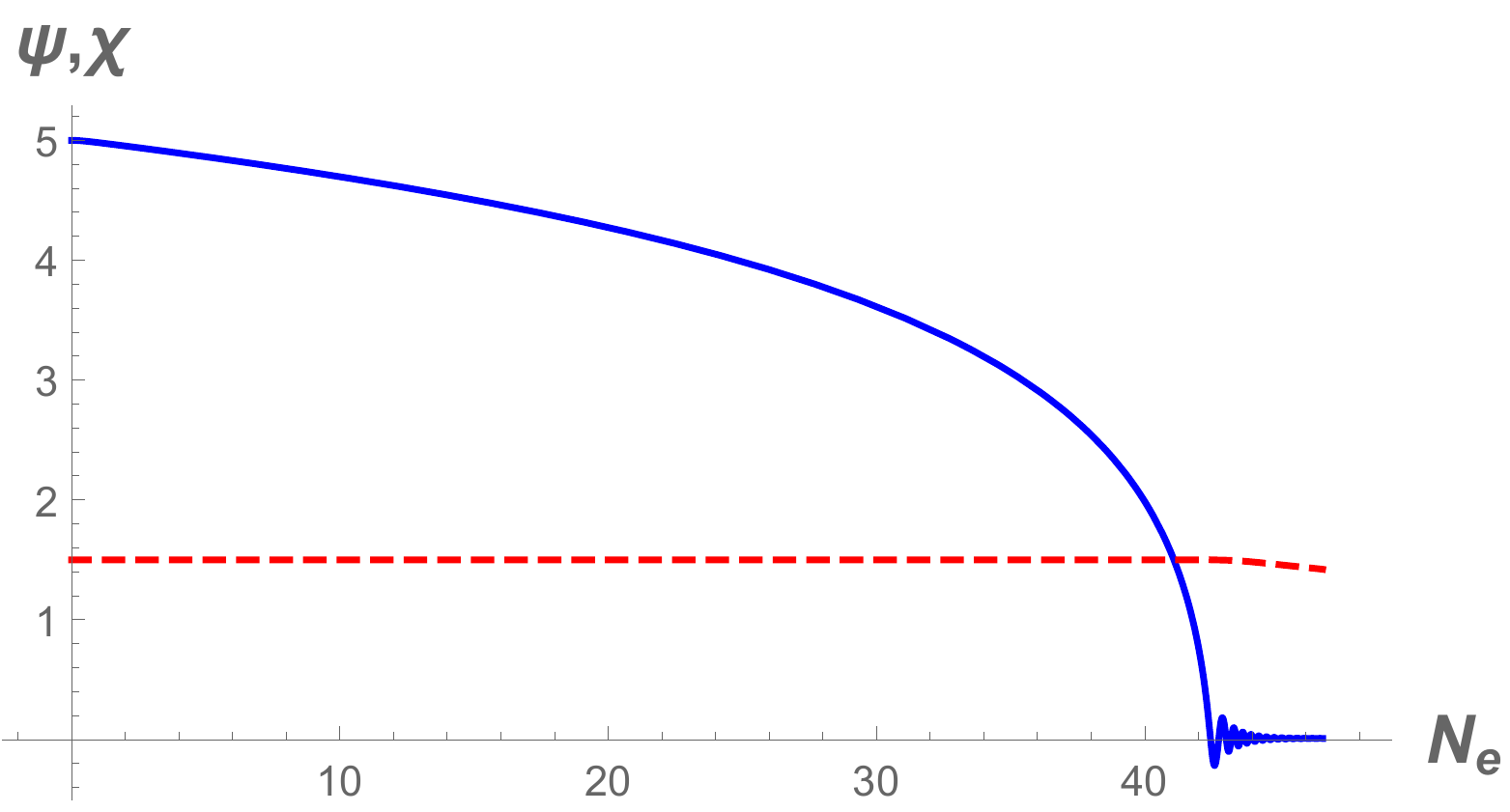}
   \caption{mass ratio=100}
 \end{subfigure}
\caption{The blue(thick) and red(dashed) curves depict the evolution of $\psi$ and  $\chi$, respectively. The free parameters are chosen as $\lambda=2\times10^{12}$ and $V_0=10^{-12}$.  The horizontal axis, $N_e$, is the number of e-folds. }
\label{fig1}
\end{figure*}

  We solve the equations of motion, (\ref{eqm1}) and (\ref{eqm2}) together with  Einstein's field equations, (\ref{H}) and (\ref{dotH}) to arrive at the evolution of the fields which are plotted in FIG.\ref{fig1}.
 To satisfy the constraint on the maximum length of the throat\cite{Mcallister}, we choose the initial value of $\chi$ equals 1.5  (which is less than the initial value of $\psi$). At the end of inflation $\chi$ decreases to a small value( from brane-inflation viewpoints, branes and anti-branes annihilate near the bottom of the throat). We define the mass ratio parameter, $\beta$, as $\beta=\frac{m_\psi}{m_\chi}$. These figures show that when $\beta$ becomes much larger than one, the scalaron traps at its minimum and the energy density of the DBI field overcome the energy density of scalaron; thus DBI field governs the dynamics. The effect of scalaron is hidden in the boost factor; $\psi$ provides enough e-folds and keep the boost factor around 1, which allows us to use slow-roll approximation and also assume that the  DBI field is potential-dominated.

\subsection{Background}\label{background}
The scalaron rolls down in the effective potential to go to its minimum, where it is trapped. The effective potential depends on both fields. It is written as follows,
 \ba
 U_{eff}=\frac{1}{8\kappa^2\mu}e^{-4\alpha\psi}{\left(e^{2\alpha\psi}-1\right)^2}-\frac{1}{4}e^{-4\alpha\psi}T^b_{DBI}.
 \ea
The extremum  value at $\psi_{min}$  satisfies ,
 \ba\nonumber
 [\frac{\alpha}{2\kappa^2\mu}e^{-4\alpha\psi}\left(e^{2\alpha\psi}-1\right)+\alpha e^{-4\alpha\psi}T_{DBI}^b]\mid_{\psi_{min}}=0,
 \ea
 solving the above equation gives,
 \ba\label{psi min}
 \psi_{min}=\frac{1}{2\alpha}\ln{\left(1-2\kappa^2\mu T_{DBI}^b\right)},
 \ea
 the condition for having a minimum ($d^2U_{eff}/d\psi^2\mid_{\psi_{min}}>0$) is always satisfied because we have,
  \ba
 \frac{d^2U_{eff}}{d\psi^2}\mid_{\psi_{min}}=\frac{\alpha^2}{\kappa^2\mu}e^{-2\alpha\psi_{min}}>0.
  \ea
 We assume that the fields are potential dominated  i.e.  $T^b_{DBI}\simeq-4U\left(\chi\right)$ and $e^{2\alpha\psi_{min}}\approx1+8\kappa^2\mu U\left(\chi\right)$ , the Friedmann equations can be approximated as,
  \ba\label{approx}
   3H^2&\simeq&\frac{1}{8k^2\mu}e^{-4\alpha\psi}[(e^{2\alpha\psi}-1)^2+8k^2\mu U(\chi)]\mid_{\psi_{min}},\label{H min}\\\nonumber
       &\simeq&e^{-2\alpha\psi}U(\chi)\\\nonumber
   -2\dot H&\simeq&e^{-2\alpha\psi}\gamma\dot{\chi}^2.\label{dot H min}
  \ea
 From now on, we dropped the index min. As mentioned before, when $\psi$ is trapped at its minimum, the dynamics is controlled by $\chi$. Comparing with usual DBI in the general relativity context shows that the effect of $\psi$  or equivalently $R^2$  term appears in $e^{-2\alpha\psi}$ factor.
  To arrive at the above equations, we assumed $\dot\psi^2\ll e^{-2\alpha\psi}\gamma\dot\chi^2$ in (\ref{dotH}). This assumption is equivalent to,
  \ba\gamma\gg\frac{4\kappa^2}{9\alpha^2\beta^2}\frac{\chi^2}{1+\frac{2\kappa}{3\beta}\chi^2}.\ea
 This condition is satisfied when $\beta\gg1$,  i.e. when there is a heavy scalaron. In the following, we assume this condition is satisfied.
  Differentiating (\ref{psi min}) with respect to time gives the change of the minimum of $\psi$, as $\chi$ evolves,
  \ba\label{psi min diff}
  \dot{\psi}=\frac{4\kappa^2\mu}{\alpha}e^{-2\alpha\psi}U_{,\chi}\dot{\chi}.
  \ea
   The coupling between the two fields causes Hubble friction term and potential terms  dominate in the DBI equation of motion (\ref{eqm2}), then we have
  \ba\label{app  dot chi}
  \dot\chi\simeq-e^{-2\alpha\psi}\frac{U_{,\chi}}{3H\gamma}
  \ea
  We  also define the slow-roll parameters as usual,
\ba
\epsilon\equiv-\frac{\dot{H}}{H^2}=\frac{3}{2}\frac{\dot{\psi}^2+\gamma e^{-2\alpha\psi}\dot{\chi}^2}{\rho_{DBI}+\frac{1}{2}\dot{\psi}^2+w\left(\psi\right)},
\ea
Using (\ref{approx})we arrive at,
  \ba\label{epsilon}
  \epsilon&\approx&\frac{3}{2}\frac{\gamma\dot{\chi}^2}{U\left(\chi\right)}\\\nonumber
  &\approx&\frac{1}{2}e^{-2\alpha\psi}(\frac{U_{,\chi}}{U})^2.
  \ea
Similar to previous results the only difference with usual DBI model is $e^{-2\alpha\psi}$ factor. As usual we have $\ddot{a}/a=H^2\left(1-\epsilon\right)$.
The inflation ends when $\epsilon$ gets larger than 1. In our numerical analysis, we get around 55 e-folds.
  Differentiate (\ref{epsilon}) with respect to time, we arrive at the rate of change of this slow-roll parameter,
  \ba\label{dotepsilon}
  \frac{\dot\epsilon}{2H\epsilon}\simeq\frac{1}{2}s-\delta+2\left(1+12\kappa^2\mu U\left(\chi\right)\right)\epsilon,
  \ea
  with
  \ba\label{slow param}
  s=-\frac{\dot\gamma}{H\gamma}\hspace{1.5cm}and\hspace{1.5cm}\delta=\frac{1}{\gamma}\frac{U_{,\chi\chi}}{U}.
  \ea
   where "s" measures the rate of change of the sound speed and $\delta$ is equivalent to $\eta$ parameter. Note that both of these parameters has implicit dependence on $\psi$ through $e^{-2\alpha\psi}$ factor in $\gamma$ .
   From a mathematical point of view, our model is equivalent to a scalar-tensor theory\cite{ Bruck2010, Bruck2011}. But the physics behind these models is different. In our case, the canonical scalar field originated from higher-order gravity and quantum corrections rather than put inside the theory by hand.

It is worth mentioning that there are other models which have interesting motivations for deriving inflation, for example in [44], it is shown explicitly that the quantum potential plays the role of the cosmological constant and also produces the exponential expansion.

\section{Perturbations evolution and cosmological parameters}\label{perturbation}
First, we consider the evolution of linear perturbation of this model. We perturb the action (\ref{full action}) in a standard way by decomposition of the fields $\psi$ and $\chi$ into a homogeneous and perturbed part,
\ba
 \psi\left(t,\mathbf{x}\right)=\psi\left(t\right)+\delta\psi\left(t,\mathbf{x}\right)\hspace{2cm} \chi\left(t,\mathbf{x}\right)=\chi\left(t\right)+\delta\chi\left(t,\mathbf{x}\right).
\ea
The field perturbations are of linear order. We shall work in Fourier space in which the spatial derivative,$\partial$, can be replaced by $-ik$. Assume that the anisotropic stress is absent, in longitudinal gauge, the scalar perturbation of the flat FRW metric is expressed as below,
\ba
 ds^2=-\left(1+2\Phi\right)dt^2+a^2\left(t\right)\left(1-2\Phi\right)\delta_{ij}dx^{i}dx^{j}.
\ea

The equations of field perturbation are as follows
\ba\label{perturbeqm1}\nonumber
&\ddot{\delta\psi}&+3H\dot{\delta\psi}-4\dot{\Phi}\dot{\psi}+\alpha\dot{\delta\chi}\dot{\chi}e^{-2\alpha\psi}
(3\gamma-\gamma^3)\\\nonumber
&+&\delta\psi(\frac{k^2}{a^2}+w_{,\psi\psi}-4\alpha^2e^{-4\alpha\psi}[f^{-1}(4-3\gamma^{-1}-\gamma)-4U(\chi)]+
\alpha^2e^{-2\alpha\psi}(3\gamma-\gamma^3)\dot{\chi}^2)\\\nonumber
&+&\delta\chi(-4\alpha e^{-4\alpha\psi}U_{,\chi}-\alpha e^{-4\alpha\psi}\frac{f_{,\chi}}{2f^2}(8-3\gamma^{-1}-6\gamma+\gamma^3))\\\nonumber
&+&2\Phi(w_{,\psi}+\alpha e^{-4\alpha\psi}[f^{-1}(4-3\gamma^{-1}-\gamma)-4U(\chi)]+\alpha e^{-2\alpha\psi}(3\gamma-\gamma^3)\dot{\chi}^2)=0,\ea
and
\ba\label{perturbeqm2}\nonumber
&\ddot{\delta\chi}&+(3H+3\frac{\dot\gamma}{\gamma}-2\alpha\dot{\psi})\dot{\delta\chi}-\dot{\Phi}\dot{\chi}(1+3\gamma^{-2})
+\alpha\dot{\delta\psi}\dot{\chi}(1-3\gamma^{-2})\\\nonumber
&+&\delta\chi\{\gamma^{-2}\frac{k^2}{a^2}+\gamma^{-3}U_{,\chi\chi}e^{-2\alpha\psi}+\frac{f_{,\chi}}{f}\frac{\dot{\chi}\dot{\gamma}}{\gamma}-
\frac{1}{2}U_{,\chi}f_{,\chi}\gamma^{-1}\dot{\chi}^2\\\nonumber
&+&\frac{1}{2}e^{-2\alpha\psi}(1-\gamma^{-1})^2\gamma^{-2}
[\gamma(\frac{f_{,\chi}}{f^2})_{,\chi}+(\frac{f_{,\chi}}{f})_{,\chi}\frac{1}{f}(1+\gamma^{-1})\gamma^2]\}\\\nonumber
&-&\alpha\delta\psi(\gamma^{-1}(1+\gamma^{-2})U_{,\chi}e^{-2\alpha\psi}+\frac{f_{,\chi}}{f^2}\gamma^{-1}(1-\gamma^{-1})^2e^{2\alpha\psi}
-2\dot{\chi}\frac{\dot{\gamma}}{\gamma})\\\nonumber
&+&\Phi(e^{-2\alpha\psi}\gamma^{-1}(1+\gamma^{-2})U_{,\chi}
-2\dot{\chi}\frac{\dot{\gamma}}{\gamma}+\frac{f_{,\chi}}{f^2}e^{-2\alpha\psi}\gamma^{-1}(1-\gamma^{-1})^2)=0.
\ea
It is convenient to introduce gauge-invariant quantity, so-called Sasaki-Mokhanuv variables\cite{Bruck2010,Lalak2007},
\ba
 Q_\psi\equiv\delta\psi+\frac{\dot{\psi}}{H}\hspace{2cm} Q_\chi\equiv\delta\chi+\frac{\dot{\chi}}{H},
\ea
which are the scalar field perturbations in the flat gauge. In terms of these new variables the equations form a closed system,
\ba
\ddot{Q_\psi}&+&3H\dot{Q_\psi}+B_\psi\dot{Q_\chi}+\left(\frac{k^2}{a^2}+C_{\psi\psi}\right)Q_\psi+C_{\psi\chi}Q_\chi=0,\\
\ddot{Q_\chi}&+&\left(3H-2\alpha\dot{\psi}+3\frac{\dot{\gamma}}{\gamma}\right)\dot{Q_\chi}\\\nonumber
&+&B_\chi\dot{Q_\psi}\left(\frac{k^2}{a^2\gamma^2}+C_{\chi\chi}\right)Q_\chi+C_{\chi\psi}Q_\psi=0.
\ea
 with the coefficients as
\ba\nonumber
B_\chi&=& -\alpha(\frac{3}{\gamma^2}-1)\dot{\chi}-\frac{\dot{\psi}\dot{\chi}}{2H}(1 - \frac{1}{\gamma^2}),\\\nonumber
B_\psi&=&-e^{-2\alpha\psi}\gamma^3B_\chi,\\\nonumber
C_{\psi\psi}&=&-\alpha\frac{\dot{\psi}}{H} e^{-4\alpha\psi}f^{-1}(\frac{3}{\gamma} + 1)(1 -\gamma)^3 -\alpha^2 e^{-4\alpha\psi}f^{-1}(16 - 8\gamma - \frac{9}{\gamma}+\gamma^3) + 3\dot{\psi}^2 \\\nonumber
&-&\gamma^3(1 + \frac{1}{\gamma^2}) e^{-2\alpha\psi}\frac{\dot{\psi}^2\dot{\chi}^2}{4H^2}-\frac{\dot{\psi^4}}{2H^2}+\alpha e^{-4\alpha\psi}\frac{2\dot{\psi}}{H}(\frac{1}{2\kappa\mu}(1-e^{2\alpha\psi})-4U(\chi))\\\nonumber&+&\alpha^2e^{-4\alpha\psi}
\left(\frac{1}{\kappa^2\mu}(2-e^{2\alpha\psi})+16U(\chi)\right),\\\nonumber
C_{\psi\chi}&=&\frac{e^{-4\alpha\psi}\dot{\psi}}{4H}\frac{f_{,\chi}}{f^2}\frac{1}{\gamma}(1 -\gamma)^2(\gamma^2 + 2\gamma- 1) + 3\gamma e^{-2\alpha\psi}\dot{\psi}\dot{\chi}-\gamma^4(1 + \frac{1}{\gamma^2})e^{-4\alpha\psi}\frac{\dot{\psi}\dot{\chi}^3}{4H^2}\\\nonumber
&+&\frac{1}{2}\alpha e^{-4\alpha\psi}f^{-1}(\frac{3}{\gamma}+ 1)(1 - \gamma)^3\left(\frac{f_{,\chi}}{f}-\frac{e^{-2\alpha\psi}\dot{\chi}\gamma}{H}\right)-\gamma e^{-2\alpha\psi}\frac{\dot{\psi}^3\dot{\chi}}{2H^2}+e^{-4\alpha\psi}\frac{\dot{\psi}}{H}U(\chi)_{,\chi}\\\nonumber
&+&\alpha\gamma e^{-6\alpha\psi}\frac{\dot{\chi}}{H} (\frac{1}{2\kappa^2\mu}(1-e^{2\alpha\psi}-1)-4U(\chi))-4\alpha e^{-4\alpha\psi}U_{,\chi},\\\nonumber
C_{\chi\chi}&=&e^{-4\alpha\psi}\frac{\dot{\chi}}{H}\frac{f_{,\chi}}{f^2}(1-\frac{1}{\gamma})^2-(\frac{f_{,\chi}}{f}+
\frac{e^{-2\alpha\psi}\dot{\chi}\gamma}{H})\frac{\dot{\gamma}}{\gamma}\dot{\chi}\\\nonumber
&-&\frac{1}{2\gamma}f_{,\chi}\dot{\chi}^2U_{,\chi}+\frac{1}{2}e^{-2\alpha\psi}
(1-\frac{1}{\gamma})^2[\frac{1}{\gamma}(\frac{f_{,\chi}}{f^2})_{,\chi}+(1+\frac{1}{\gamma})f^{-1}(\frac{f_{,\chi}}{f}t)_{,\chi}]\\\nonumber
&+&\frac{3}{2}e^{-2\alpha\psi}\dot{\chi}^2\gamma(1+\frac{1}{\gamma^2})-e^{-4\alpha\psi}\gamma^2\frac{\dot{\chi}^4}{2H^2}-
e^{-2\alpha\psi}\gamma(1+\frac{1}{\gamma^2})\frac{\dot{\chi}^2\dot{\psi}^2}{4H^2}\\\nonumber
&+&e^{-4\alpha\psi}\frac{\dot{\chi}}{H}(1+\frac{1}{\gamma^2})U_{,\chi}+\frac{1}{\gamma^3}e^{-2\alpha\psi}U_{,\chi\chi},\\\nonumber
\ea
\ba\nonumber
C_{\chi\psi}&=&(-2e^{-2\alpha}+\frac{\dot{\psi}}{H})(\frac{1}{2}e^{-2\alpha\psi}\frac{f_{,\chi}}{f^2\gamma}(1-\frac{1}{\gamma})^2
-\frac{\dot\gamma}{\gamma}\dot\chi)+2\alpha\frac{e^{-4\alpha\psi}\dot{\chi}}{H}f^{-1}(1-\frac{1}{\gamma})^2
\\\nonumber
&-&\gamma\frac{e^{-2\alpha\psi}\dot\psi\dot\chi^3}{2H^2}+\frac{1}{2}(1+\frac{1}{\gamma^2})\\\nonumber
&&\left(3\dot\psi\dot\chi-\frac{\dot\phi^3\dot\chi}{2H^2}-\frac{2\alpha}{\gamma}e^{-2\alpha\psi} U_{,\chi}+\frac{\dot\psi e^{-2\alpha\psi}U_{,\chi}}{\gamma H}+\alpha e^{-4\alpha\psi}(\frac{1}{2k^2\mu}(e^{2\alpha\psi}-1)-4U(\chi))\frac{\dot\chi}{H}\right).\nonumber
\ea

Similar to single-field perturbation analysis in canonical  and DBI models, we introduce two auxiliary fields as,
\ba
u_\psi=a Q_\psi,\hspace{2cm}u_\chi=ae^{-\alpha\psi}c_s^{-3/2} Q_\chi.
\ea
The equations of motion in terms of conformal time can be rewritten in a more symmetric form,
\ba
u''_\psi&-&Bu'_\chi+[k^2+a^2C_{\psi\psi}-\frac{r''_\psi}{r_\psi}]u_\psi+[\frac{r_\psi}{r_\chi}a^2C_{\psi\chi}+B\frac{r'_\chi}{r_\chi}]u_\chi=0\label{upsi}\\
u''_\chi&+&Bu'_\psi+[k^2c_s^2+a^2C_{\chi\chi}-\frac{r''_\chi}{r_\chi}]u_\chi+[\frac{r_\chi}{r_\psi}a^2C_{\chi\psi}-B\frac{r'_\psi}{r_\psi}]u_\psi=0\label{uchi}
\ea
 where $()'$ denotes the derivative with respect to conformal time and we define $c_s=1/\gamma$, $r_\chi=ae^{-\alpha\psi}\gamma^{3/2}$, $r_\psi=a$, and $B=r_\chi B_\chi$.
The co-moving curvature perturbation,
can be express in terms of gauge invariant variables $Q_\psi$ and $Q_\chi$  in a simple form\cite{Bruck2010},
\ba \label{R}\mathcal{R}=\frac{H}{-2\dot{H}}[\dot\psi Q_\psi+e^{-2\alpha\psi}\gamma\dot\chi Q_\chi].\ea

\subsubsection*{The evolution of perturbations for a trapped scalaron:}
\begin{figure}[H]
\includegraphics[keepaspectratio=true,scale=0.6]{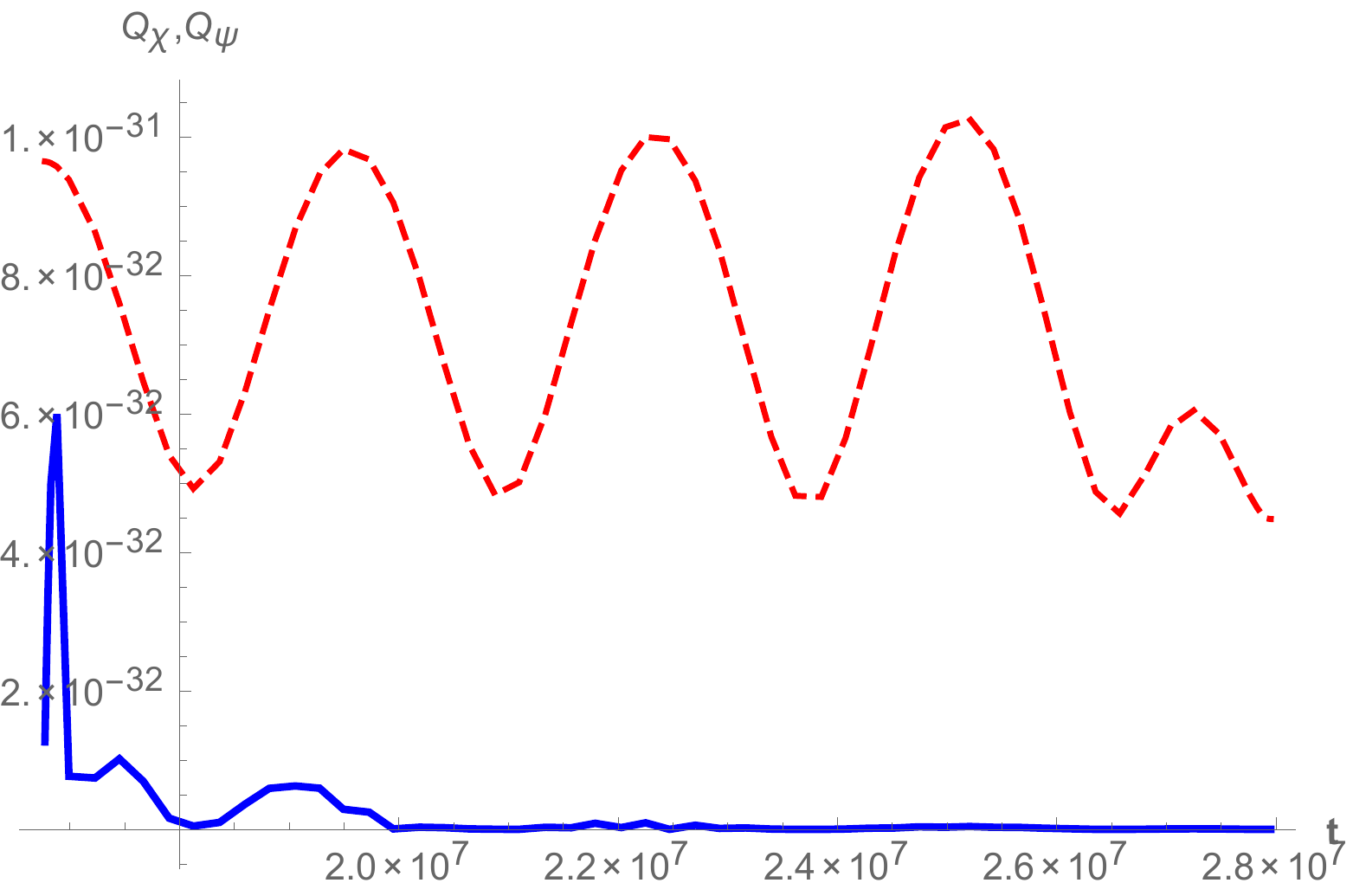}
\caption{We depict  the contribution of the scalaron (blue thick curve) and DBI field (red  dashed curve) in curvature perturbation (\ref{R}),  after $\psi$ trapped at the minimum of the effective potential. The  horizontal axis is time. }
\label{fig2}
\end{figure}

The contribution of $Q_\psi$ in curvature perturbation can be ignored when the scalaron,  $\psi$,  traps in the minimum of the effective potential. In this case, It is possible to treat  the system of equations as a single field DBI model with modified boost factor. Numerical analysis supports this approximation\footnote{In our numerical code we got some help from numerical code mTransport\cite{numeric}}(Fig(\ref{fig2})).
After that, the dynamics is governed by the DBI field.

 Therefore, the perturbation equations (\ref{upsi} and \ref{uchi}) are estimated as follows,
\ba
u''_\chi+[k^2c_s^2+a^2C_{\chi\chi}-\frac{r''_\chi}{r_\chi}]u_\chi\simeq0.
\ea
  insertion of (\ref{app  dot chi}) into (\ref{slow param}), gives $C_{\chi\chi}$ and the derivative of  $r_\chi$  in terms of slow-roll parameters (up to the first order) as
 \ba\label{c chi chi}
a^2 C_{\chi\chi}&\simeq&3\mathcal{H}^2[\delta-s-2\epsilon+8\kappa^2\mu U\epsilon\left(1-\gamma^{-2}\right)],
 \ea
 and
 \ba\label{r''}
 \frac{r''_\chi}{r_\chi}&\simeq&\mathcal{H}^2\left(2-\frac{2}{9}s-\left(1-24\kappa^2\mu U\right)\epsilon\right)
 \ea
where $\mathcal{H}=a'/a$ and $\mathcal{H}'=\mathcal{H}^2\left(1-\epsilon\right)$. The background variable z is defined as usual,
\ba z\equiv\frac{a\gamma\sqrt{\rho+p}}{H}=a\gamma\sqrt{2\epsilon}\ea
where we used the fact $-2\dot H=\rho+p$. Combination of (\ref{c chi chi}) and (\ref{r''}) gives,
\ba
a^2C_{\chi\chi}-\frac{r''_\chi}{r_\chi}\simeq-\frac{z''}{z}+24\kappa^2\mu U\epsilon\mathcal{H}^2\left(1+c_s^2\right).\ea
 The first term is almost the same as single field k-inflation\cite{mukhanov1},  in which $\frac{u_\chi}{z}$ is constant for small k; the second term is a small correction of order $\epsilon$ which is proportional to $\left(1+c_s^2\right)$. At lowest order, we ignore the second term;
 \ba\label{u} u''_\chi+\left(k^2c_s^2-\mathcal{H}^2[2-\frac{3}{2}s-3\delta+\left(5+72\kappa^2\mu U\right)\epsilon]\right)u_\chi\simeq0.\ea

 Ignoring the perturbation of $\psi$ in the co-moving curvature perturbation (\ref{R}) gives,
 \ba \mathcal{R}\simeq\frac{e^{-\alpha\psi}\gamma^{1/2}}{\sqrt{2\epsilon}}Q_\chi=\frac{u_\chi}{z}.\ea
 The power spectrum is as
 \ba\label{powersp}\mathcal{P}_{\mathcal{R}}\simeq\frac{k^3}{2\pi^2}|\frac{u_\chi}{z}|^2.\ea

 For solving eq.(\ref{u}) we follow the approach in \cite{Bruck2011}, define a new time variable as,
 \ba\label{y}y\equiv\frac{c_s k}{a H}=\frac{c_s k}{\mathcal{H}}.\ea
 With this definition at sound horizon crossing,  we have  $y=1$.
  The derivatives of $u_\chi$ can be expressed in terms of slow-roll parameters,
 \ba\nonumber u'_\chi=-c_s k\left(1-\epsilon-s\right)\frac{du_\chi}{d\tau},\ea
  and
  \ba\nonumber u''_\chi=\mathcal{H}^2[\left(1-\epsilon-s\right)^2y^2\frac{du_\chi^2}{dy^2}-s\left(1-\epsilon-s\right)y\frac{du_\chi}{dy}],\ea
  where we have used  $\mathcal{H}'=\mathcal{H}^2\left(1-\epsilon\right)$. Substituting in (\ref{u}) gives,
  \ba\label{uy}\nonumber
   y^2\frac{d^2u_\chi}{dy^2}+(1-2p) y\frac{du_\chi}{dy}+\left(l^2y^2+p^2-\nu^2\right)u_\chi=0,
   \ea
  with
  \ba
  p&=&\frac{1}{2}(1+s),\\
  l&=&(1-\epsilon-s)^{-1},\\
  \nu&=&\frac{3}{2}+s-\delta+3\epsilon(1+8\kappa^2\mu U.)
  \ea

The solution of (\ref{uy}) is of the form $u_\chi=y^pJ_\nu(ly)$. $J_\nu$  denotes Bessel function of order $\nu$. Instead of Bessel functions, we write the solution in terms of Hankel functions, which are more appropriate for our purpose. In the short wavelength limit , ($y\gg1$), the solution is given by positive frequency mode,
$\frac{1}{\sqrt{2c_sk}}e^{-ic_sk\tau}$, where $\tau$ is conformal time.
Only $H^{(1)}_\nu( ly)$ can satisfy this initial condition; therefore, the solution is
\ba u_\chi( y)=\frac{1}{2}\sqrt{\frac{\pi}{c_sk}}\sqrt{\frac{y}{1-\epsilon-s}}H^{(1)}_\nu(\frac{y}{1-\epsilon-s}).\ea
In the long-wavelength limit ($y\ll1$) we have,
$H^{(1)}_\nu(ly)\sim\sqrt{\frac{2}{\pi}}e^{-i\pi/2}2^{\nu-\frac{3}{2}}\frac{\Gamma(\nu)}{\Gamma(3/2)}y^{-\nu}$; so,  the solution is
\ba |u_\chi|\sim2^{\nu-\frac{3}{2}}\frac{\Gamma(\nu)}{\Gamma(3/2)}(1-\epsilon-s)^{\nu-\frac{1}{2}}\frac{y^{\nu-\frac{1}{2}}}{\sqrt{2c_sk}}.\ea
Replacing in (\ref{powersp}) we arrive at
\ba\mathcal{P}_{\mathcal{R}}^{1/2}\simeq(\frac{\mathcal{V}(\nu)}{\pi})\frac{H}{\sqrt{c_s\epsilon}}y^{\frac{3}{2}-\nu},\ea
with $\mathcal{V}\equiv2^{\nu-3}(1-\epsilon-s)^{\nu-\frac{1}{2}}\Gamma(\nu)/\Gamma(\frac{3}{2})$.

 It can be shown that $\frac{d}{dy}(\frac{H}{\sqrt{c_s\epsilon}}y^{\frac{3}{2}-\nu})\simeq0,$, which insures us that the power spectrum is independent of y and can be evaluated at any preferred y value\cite{Bruck2011,generalizations of slow-roll,y1,y2}, hence the sound crossing formalism is applicable.

\begin{figure}[H]
\includegraphics[keepaspectratio=true,scale=0.5]{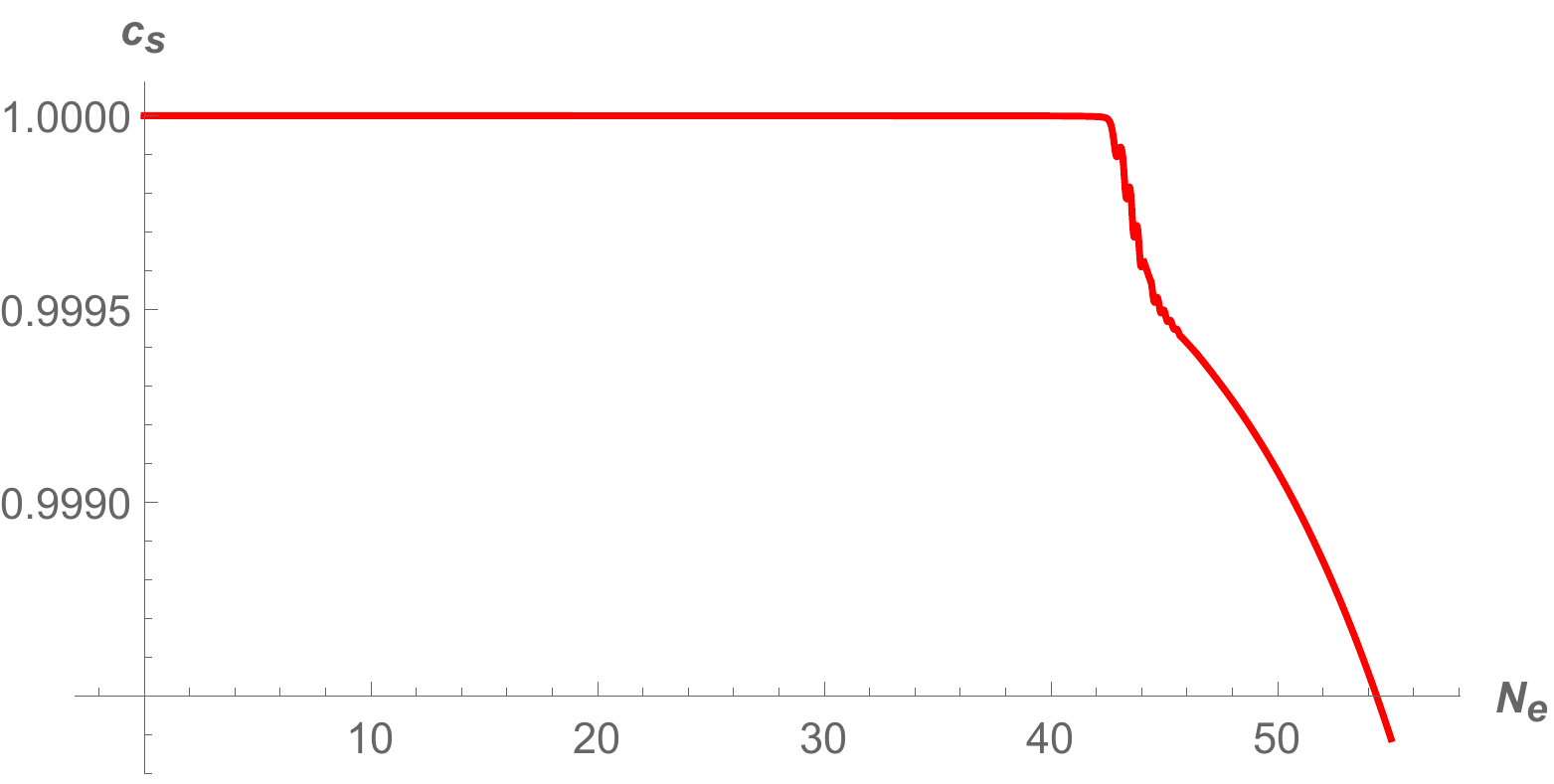}
\caption{The sound speed versus  the number of e-folds is shown. Parameters value are chosen as $\lambda=2\times10^{12}$, $V_0=10^{-12}$,  and $\beta=55$.}
\label{fig6}
\end{figure}

Using this gives the spectral index (up to first order in slow-roll parameters) as,
\ba
n_s-1&=&3-2\nu\\\nonumber
     &=&-2s+2\delta-6\epsilon(1+8\kappa^2\mu U).
\ea
Replacing the slow-roll parameters we arrive  at
\ba
n_s-1&=&2\frac{\dot\gamma}{H\gamma}-8\frac{1}{\gamma\chi^2}.
\ea
At the end of inflation, only the DBI field drives inflation, so we ignore the perturbation of scalaron. It is reasonable to assume that the results obtained in the DBI inflation are applicable to this model; for example, the tensor-to-scalar ratio must be $r\simeq16\epsilon c_s$.
The non-Gaussianity is also given by $f^{DBI}_{NL}\simeq-0.3\left(c_s^{-2}-1\right)$. To be more precise, one can apply the result of \cite{langlois2008} and \cite{koyoma2008}to this model and obtain the third-order action.
The effect of $R^2$ gravity on the DBI field keeps the sound speed close to one (see Fig.(\ref{fig6})) i.e keeps $f^{DBI}_{NL}$ very small.

\subsection{Numerical Analysis}\label{numerical analysis}
In this section, we check the compatibility of our model with Planck 2018 data.
Our analysis shows that the amount of inflation depends on the $\psi$ initial value. We pick the initial value of $\psi$ so that to obtain enough e-folds.
Motivated by brane inflation, we choose the initial value of $\chi$ around 1\cite{Mcallister}(through this work we choose 1.5). In the DBI part, we have three undetermined parameters,the mass of $\chi$ (m), $\lambda$, and $V_0$. We investigate the effect of varying these parameters in this section.
There is also another parameter in the $R^2$ part of the action, which is denoted by $\mu$. As mentioned before, we select $\mu\sim10^9$. Since the mass square of the scalaron is proportional to the inverse of $\mu$, we have $m_\psi\sim1.3\times10^{-5}$ (in natural units).

\begin{figure}[H]
  \includegraphics[keepaspectratio=true]{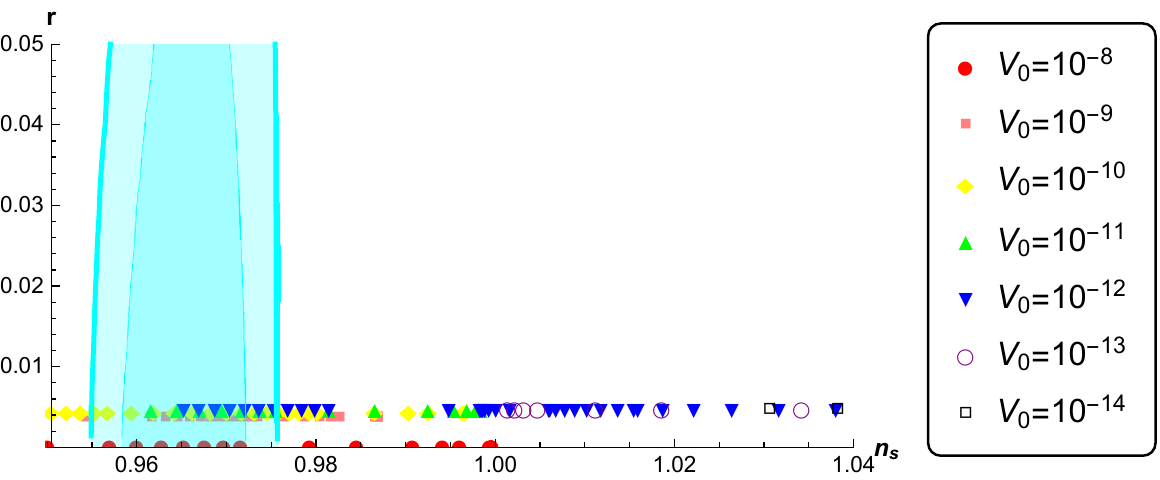}
   \caption{The tensor to scalar ratio versus the spectral index is depicted, we choose $\lambda=2\times10^{12}$. The colored regions are  $68\%$ and $95\%$  confidence level  of TT,TE,EE+lowE+lensing Planck2018 data.}
   \label{fig2a}
  \end{figure}

\begin{figure}[H]
  \includegraphics[keepaspectratio=true]{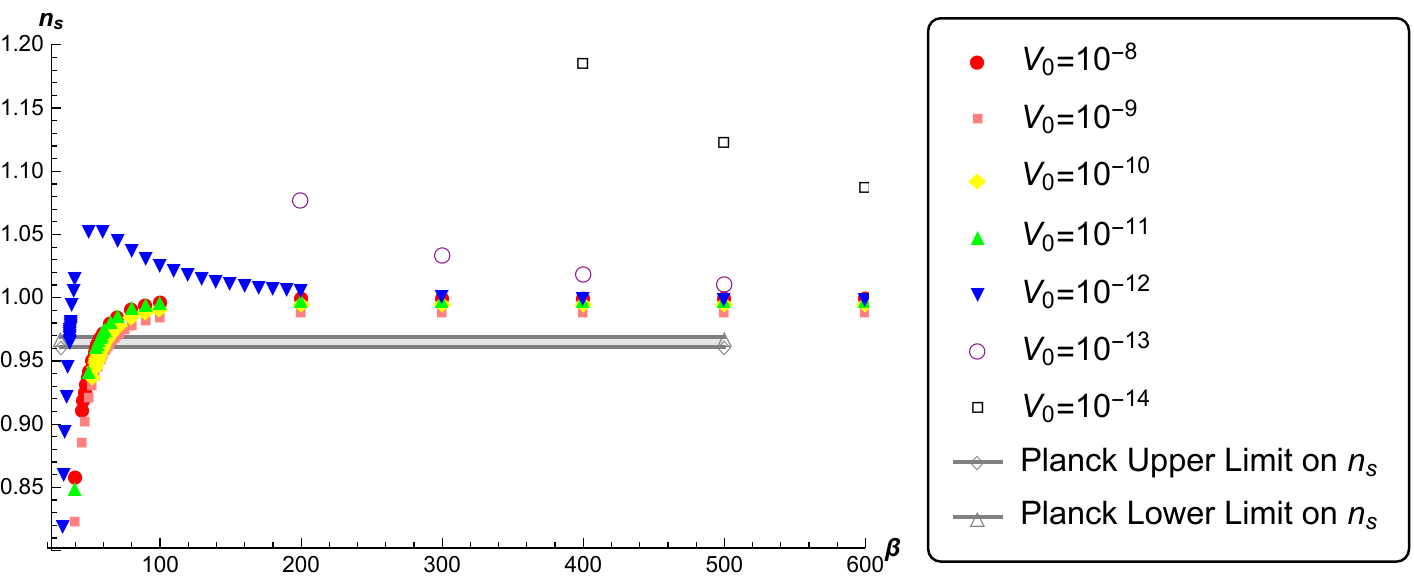}
   \caption{ We plot $n_s$ versus beta (the mass ratio), the narrow gray band shows the Planck limit. We choose $\lambda=2\times10^{12}$.}
   \label{fig2b}
 \end{figure}

\begin{figure}[H]
  \includegraphics[keepaspectratio=true]{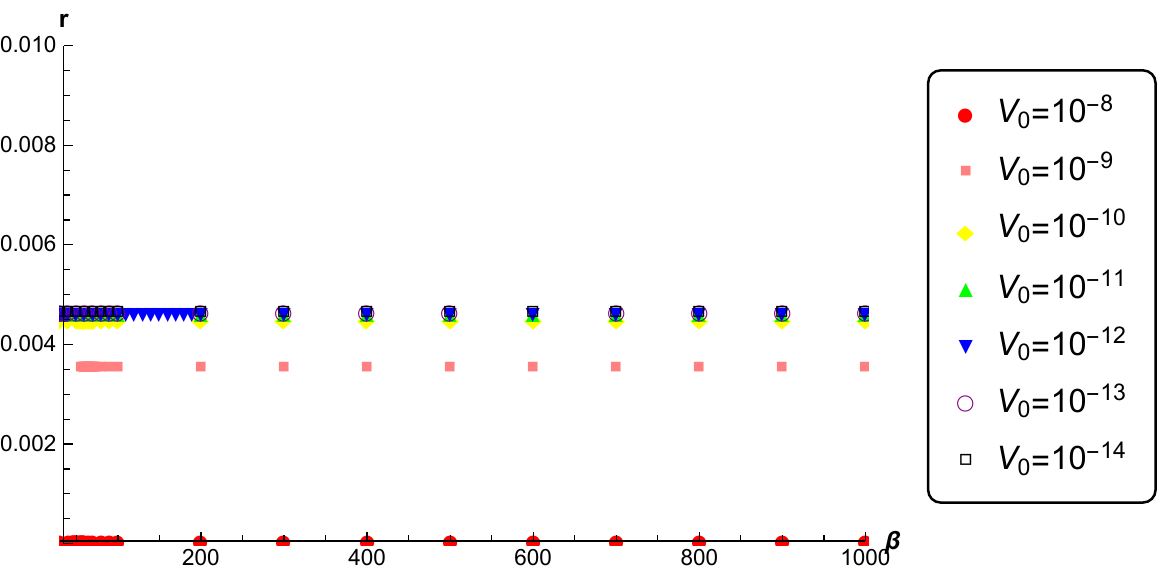}
   \caption{The tensor to scalar ratio is shown versus beta (the mass ratio). We choose $\lambda=2\times10^{12}$.
   Our results are much smaller than the Planck limit $r<0.064$}
   \label{fig2c}
 \end{figure}

We change the mass ratio parameter, $\beta$, to obtain the spectral index and the tensor-to-scalar ratio. As previously stated, in the DBI part of the action, there are also two other parameters, $\lambda$ and $V_0$.
We check the different values of these parameters. First, we inspect the different values of $V_0$; we depict the tensor-to-scalar ratio versus the spectral index in figure(\ref{fig2a}).
By increasing the mass ratio, the spectral index also increases; but the tensor-to-scalar ratio remains almost constant.
 The tensor-to-scalar ratio value is very small(in comparison with Planck upper limit 0.064). This result is similar to ordinary Starobinsky inflation\cite{Planck2018b}. To be more clear we plot spectral index (Fig(\ref{fig2b}))and the tensor-to-scalar ratio (Fig(\ref{fig2c}))with respect to mass ratio ($\beta$).


 From the above figures, it is obvious that very small and very large values of $V_0 $ are not compatible with observations. For a small value of $V_0$, the DBI potential is almost $\frac{1}{2}m^2\phi^2$. So we can conclude that this famous potential is not compatible with the Planck data, even for the DBI field. On the opposite side, for large values of $V_0$, the DBI potential is almost constant.
  It seems that to get good results, we need both parts of the potential. Therefore, we choose an intermediate value (in the other parts of this work we choose $V_0=10^{-12}$).

 \begin{figure}[H]
  \includegraphics[keepaspectratio=true]{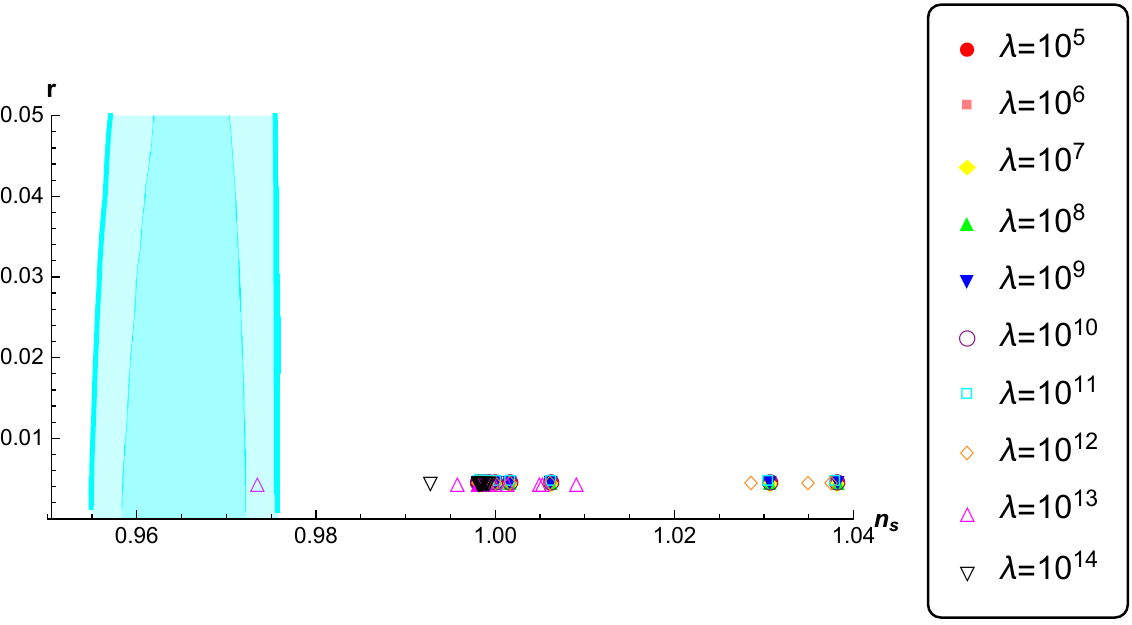}
   \caption{The tensor to scalar ratio versus the spectral index is depicted, we choose $V_0=10^{-12}$. The colored regions are  $68\%$ and $ 95\%$ confidence level of TT,TE,EE+lowE+lensing Planck2018 data.}
    \label{fig3a}
  \end{figure}
 \begin{figure}[H]
  \includegraphics[keepaspectratio=true]{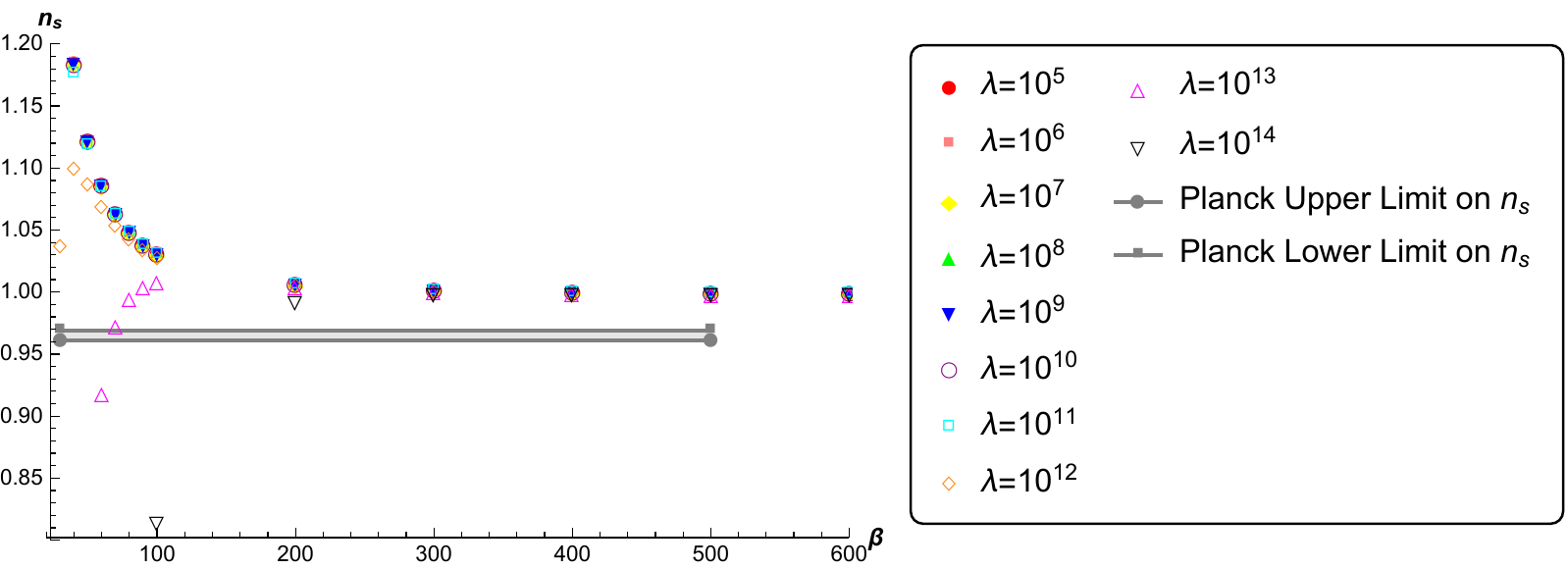}
   \caption{ We plot $n_s$ versus beta (the mass ratio), we choose $V_0=10^{-12}$. The narrow gray band shows the Planck limit.}
    \label{fig3b}
 \end{figure}
 \begin{figure}[H]
  \includegraphics[keepaspectratio=true]{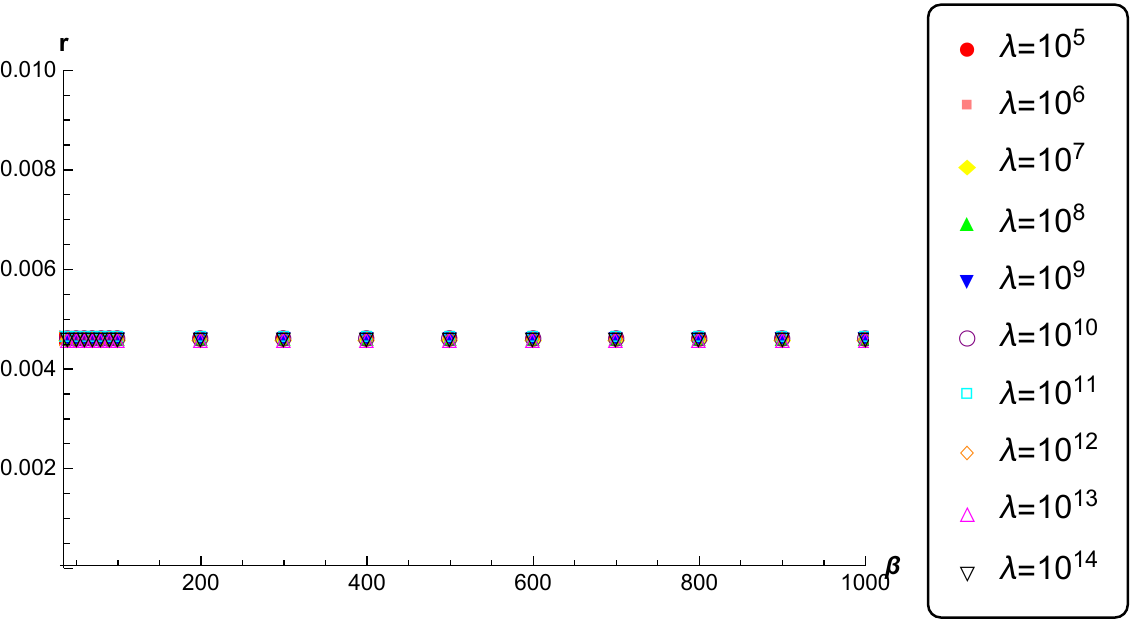}
   \caption{The tensor to scalar ratio is shown versus beta (the mass ratio), we choose $V_0=10^{-12}$. These plots are for different value of  constant part of DBI potential. Our results are much smaller than the Planck limit $r<0.064$. }
    \label{fig3c}
 \end{figure}
\begin{figure}[H]
\includegraphics[keepaspectratio=true]{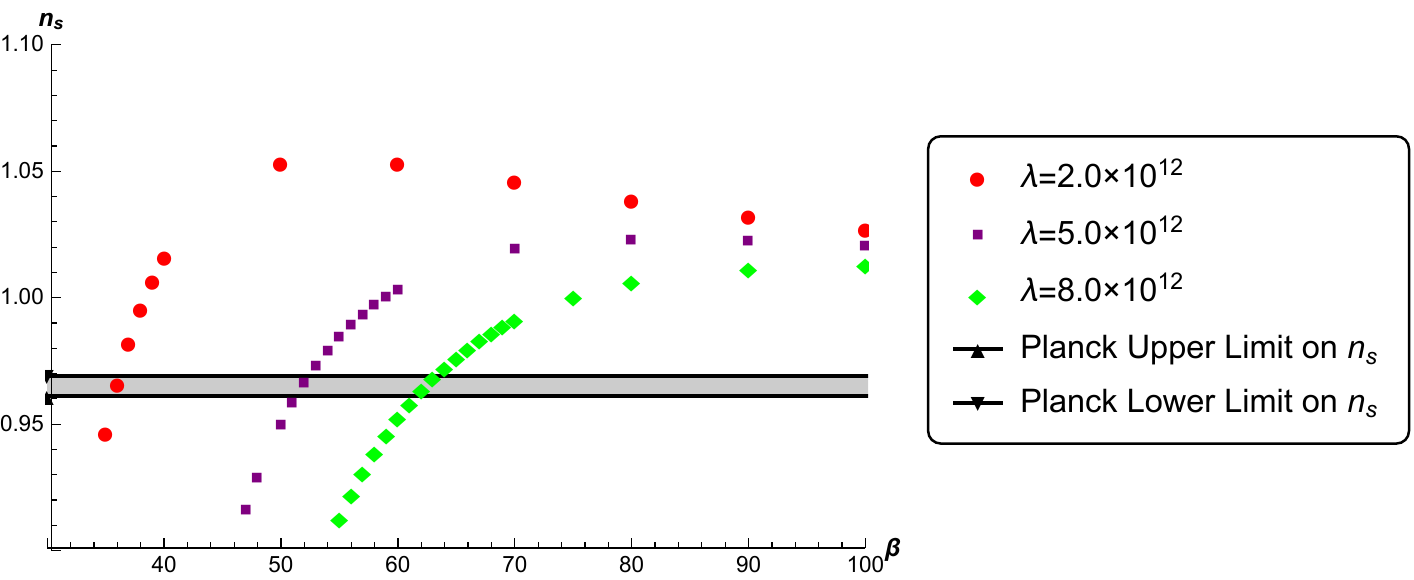}
\caption{ We plot the spectral index by varying the $\lambda$ parameter. The gray area is allowed value by Planck2018. As before $V_0=10^{-12}$ }
\label{fig4}
\end{figure}
\begin{figure}[H]
\includegraphics[keepaspectratio=true]{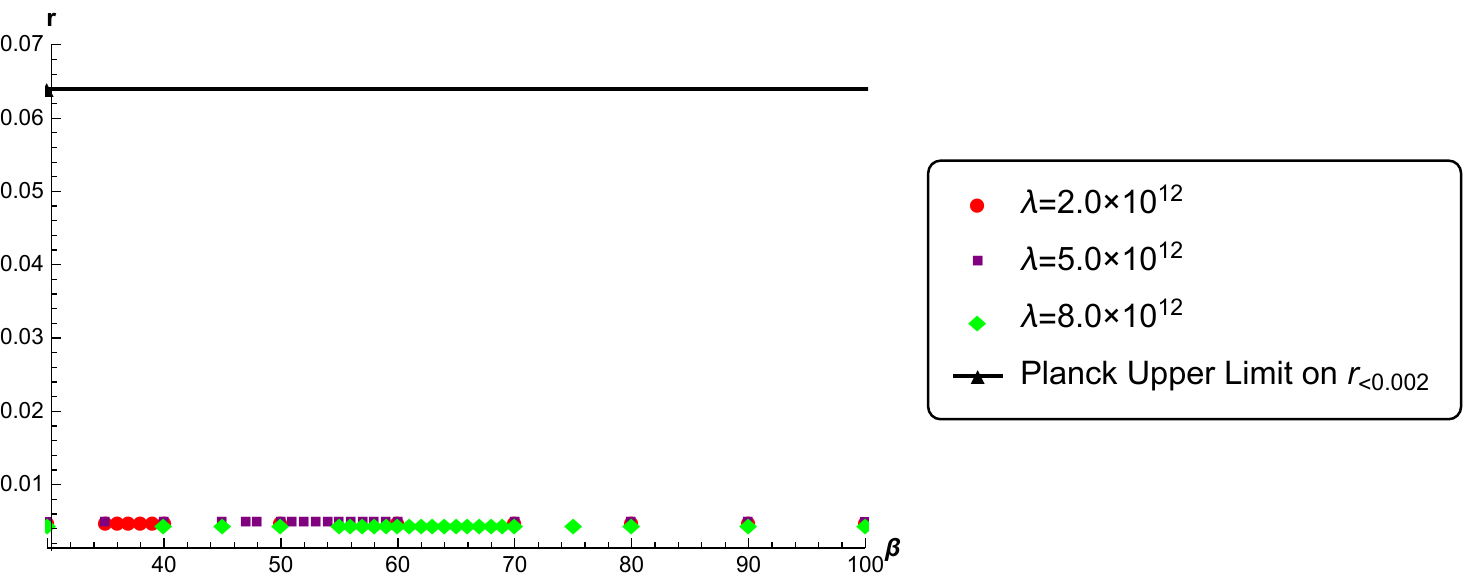}
\caption{ We plot the tensor to scalar ratio by varying the $\lambda$ parameter. The gray area is allowed value by Planck2018. As before $V_0=10^{-12}$ }
\label{fig5}
\end{figure}

To find out the effect of the other parameter, $\lambda$ we again plot r with respect $n_s$ by varying the mass ratio($\beta$) for different value of $\lambda$ (Fig(\ref{fig3a})). We also plot r (Fig(\ref{fig3b})) and $n_s$  (Fig(\ref{fig3c})) with respect to $\beta$ separately.
These figures indicate that, only intermediate values,  around $10^{12}$ to $10^{13}$ gives  compatible results, therefore for a closer look, we plot r (Fig(\ref{fig4})) and $n_s$ (Fig(\ref{fig5})) for $\lambda$ in this range.



Our analysis shows that it is possible to get the spectral index  and the tensor to scalar ratio in the  Planck range for the appropriate choice of parameters and initial conditions.

In addition, from figures \ref{fig2c}, \ref{fig3c} and \ref{fig5}
 one can conclude that in this model the scalar-tensor-ratio is very small, $r<0.01$ regardless of the spectral index. This property is different from other models, which consider extra fields in the context of F(R) gravity.


Our numerical analysis shows that regardless of the mass ratio, the sound speed is near one (see FIG\ref{fig6}. As before we set the initial values of $\chi=1.5$ and $\psi=5.3$. According to our analysis, the main results are not sensitive to initial conditions.



\section{Conclusion}\label{conclusion}
We studied the effect of the existence of a DBI field in the Starobinsky inflation, i.e. $R+cR^2$ gravity.
In this model, there are two fields: the DBI field and the scalaron. Therefore, our model is within the context of a general multi-field model.
We have shown that when the mass of the scalaron is much greater than the mass of the DBI field, the DBI field drives the inflation. In this case, the scalaron is trapped at its minimum. Before the trapping of $\psi$, the DBI field is almost constant. From the brane inflation point of view, it means that the branes move very slowly. After $\psi$ traps at its minimum, the DBI field begins to decrease i.e. the branes get closer together. Although the DBI field drives inflation, the boost factor and other quantities have implicit dependence on $\psi$. In this model, the boost factor is smaller than the single DBI model due to the existence of $e^{2\alpha\psi}$ in the square root. Hence in Starobinsky gravity the level of non-Gaussianity of DBI model decreases.
It is possible to ignore the fluctuation of scalaron when it is trapped at the minimum, so only the DBI field contributes to curvature perturbation, spectral index, tensor to scalar ratio, and other quantities that are related to the field perturbations.
Before trapping, the scalaron contribution to the energy density is much greater than the contribution of the DBI field. Therefore, the Hubble parameter, and consequently, the maximum number of e-folds, have a strong dependence on the scalaron. But due to the heaviness of scalaron, its perturbations are suppressed, and only the perturbations of the DBI field contribute to the curvature perturbation. This issue has been checked numerically.

 The main result of all these works is reducing the boost factor of the DBI field. Therefore the amount of non-Gaussianity is also decreased, which is compatible with observational data. But as stated before, our numerical results show that the spectral index, which is caused by simple potential $\frac{1}{2}m^2\phi^2$, is not compatible with Planck2018 data. To overcome this problem, we considered a well-motivated potential for the DBI part rather than a simple square potential.

 From mathematical point of view, this model is equivalent to a scalar-tensor model. In \cite{Bruck2010} and \cite{Bruck2011} DBI field in scalar-tensor theories are investigated. Our mathematical analysis is very similar to them. Our results are also
compatible with their results.Even though mathematics is the same, the physics of these models are different. In our case, the canonical field originates from quantum corrections, which are included in the $R^2$ term. We consider the brane inflation in Starobinsky gravity and investigate the effect of the existence of the DBI field in this theory.
 We have shown that with appropriate initial conditions, we get 50-60 e-folds at the end of inflation. As previously mentioned, this model is compatible with the Planck constraints on the spectral index and the tensor to scalar ratio. (see figure  (\ref{fig3a})).



\end{document}